\documentclass[12pt,preprint]{aastex}
\usepackage[]{natbib}
\usepackage{color}
\usepackage{hyperref}

\newcommand{\id}{ {\rm d} }
\newcommand{\ii}{ {\rm i} }

\begin{document}

\shorttitle{}
\shortauthors{}

\title{Solar H-$\alpha$ Oscillations from Intensity and Doppler Observations}

\author{Jason Jackiewicz}
\affil{New Mexico State University, Department of Astronomy, P.O. Box
  30001, MSC 4500, Las Cruces, NM 88003, USA}
\email{jasonj@nmsu.edu}

\author{K.S. Balasubramaniam}
\affil{Space Vehicles Directorate, Air Force Research Laboratory, Kirtland AFB, NM 87114, USA}

\begin{abstract}

%Context: 
Chromospheric wave activity around flares and filaments has been a research focus for years, and could  provide indirect measurements of  local conditions that are not otherwise accessible. One interesting observed phenomenon is oscillations in filaments, activated by distant flares and the large-scale waves they produce. Characteristics of these oscillations, such as periods, amplitudes, and lifetimes, can provide unique  information about the filament.

We measure oscillation properties in flares and filaments from H$\alpha$ chromospheric data using a new method that provides important spatial and frequency content of the dynamics. We apply the method to two flare events where filaments are observed to oscillate and determine their properties.

We find strong oscillatory signal in flaring active regions in the chromosphere over a range of frequencies. Two filaments are found to oscillate without any detectable chromospheric wave acting as an activation mechanism. We  find that filaments oscillate with periods of tens of minutes, but variations are significant at small spatial scales along the filamentary region.

The results suggest there is a frequency dependence of the oscillation amplitude, as well as a spatial dependence along single filaments that is more difficult to quantify.  It also appears that the strength of the oscillations do not necessarily depend on the strength of the trigger, although there are other possible effects that make this conclusion preliminary. Applications of this technique to other events and different data sets will provide important new insights into  the local energy densities and magnetic fields associated with dynamic chromospheric structures.

\end{abstract}


\section{Introduction}

It has been known for a long time \citep{ramsey1966} that oscillations in solar prominences and filaments can be excited by distant flares.  Such events have generally been characterized by so-called large- and small-amplitude oscillations according to the amplitude of the their observed velocities (compared to the chromospheric sound and Alfv\'{e}n velocities). Large-amplitude filament oscillations display speeds upwards of $20\,{\rm km\,s^{-1}}$ and periods of tens of minutes to hours, while small-amplitude oscillations typically are a few ${\rm km\,s^{-1}}$, with  a less well-constrained range of periods, from minutes to hours \citep{arregui2012}. Motions are found in both transverse and longitudinal directions with respect to the filament axis.

The chromosphere displays a wealth of wave phenomena. Early detections of oscillations in H$\alpha$ measurements were carried out by  \citet{elliot1969}, and  \citet{harvey1993} used several  chromospheric lines to observe the high-frequency ($5\,{\rm mHz}$) $p$ modes, as well as  5-min.~acoustic modes, both escaping from the photosphere.  However, a flare (or some other source) activating a filament oscillation is a rather rare phenomenon, quite different than the oscillations that are always present in the chromosphere. Triggers of the filament activation are thought to be  Moreton waves or EUV waves produced from remote flares \citep{eto2002,okamoto2004,bala2007,bala2010,asai2012,li2012}, or jets and subflares \citep{vrsnak2007}, or even erupting filaments themselves \citep{isobe2006}. It is likely that intrinsic properties of the chromosphere can be deduced from studying these oscillations, as filaments seem to have associated eigenfrequencies when excited. This could be used to probe their magnetic structure and topology more precisely than traditional observations.

Recent reviews \citep{tripathi2009,arregui2012} point out that much of the detailed dynamics between the interaction of the Moreton or EUV waves with the filament, the oscillatory structure (vertical/horizontal and longitudinal/transverse), the damping mechanism, and possible causes of  filament eruptions are still very open issues, as these filament oscillations have only been  observed in  about a dozen cases.

%Winking filaments \citet{ramsey1966}. Winking filaments are a common feature, whereby the structure is visible in and out of the line wings as it changes direction. 

The scope of this work is not to directly address all the issues above, but to provide a new tool for analyzing these and similar oscillatory phenomena. We have observed two oscillating filament events, one never detected  and the other previously studied  by  \citet{gilbert2008}. Standard analysis in previous studies typically involves reducing  a three-dimensional (3D) data cube  to a one-dimensional (1D) time series  by spatially averaging over pixels in the filament that appear to be oscillating, and then performing a wavelet transform in time-frequency space to ascertain the variable frequency content. The reduction in dimensions has the effect of smearing possibly important spatial information. At the same time, the result may be strongly dependent on how the averaging is carried out.

Here we describe a new technique for analyzing phenomena that have dynamic spatial and temporal scales. It is  an extension of a wavelet-type technique,  preserving the 3D nature of the input data.   Upon application, we observe properties of chromospheric oscillations that do not easily fit into the two  categories described above. We observe important differences in flaring regions and regions of oscillating filaments, indicating different responses to the trigger mechanism. We also find anomalies in oscillating filaments that have and have not interacted with a strong Moreton wave. A standard analysis is also carried out for comparison.

To introduce this  technique, we first describe the example data sets in \S~\ref{dat}. We provide an overview of the algorithm in \S~\ref{ana}, which is then applied to the data and results are shown in \S~\ref{res}. After comparisons of the results to standard methods and  a summary are provided, discussion is given in \S~\ref{dis}.

\section{Data}
\label{dat}

\begin{figure}
  \centering
  \centerline{
    \includegraphics[width=.45\textwidth,clip=true]{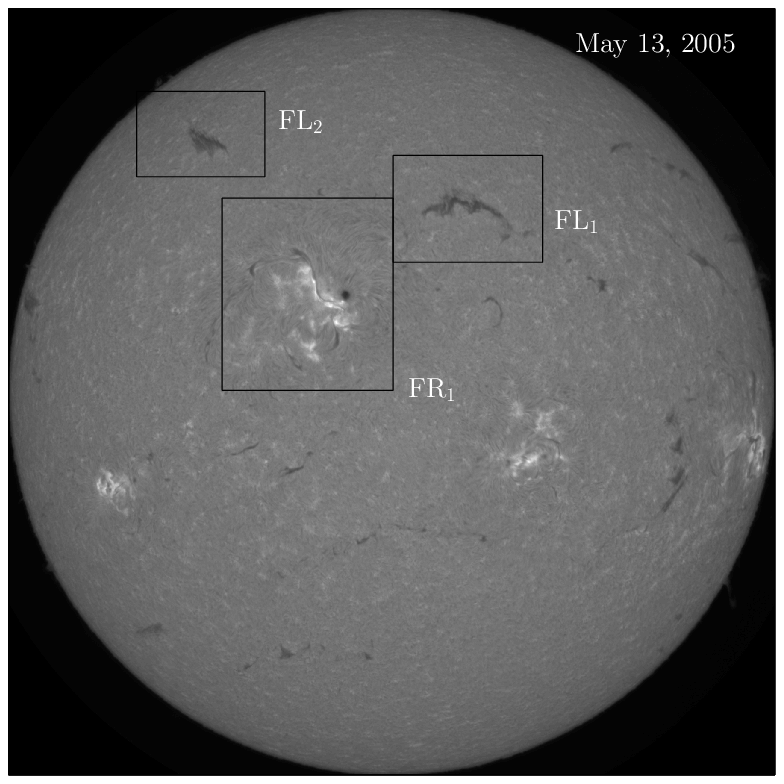} %{fd_5_2005.eps}
    \includegraphics[width=.45\textwidth,clip=true]{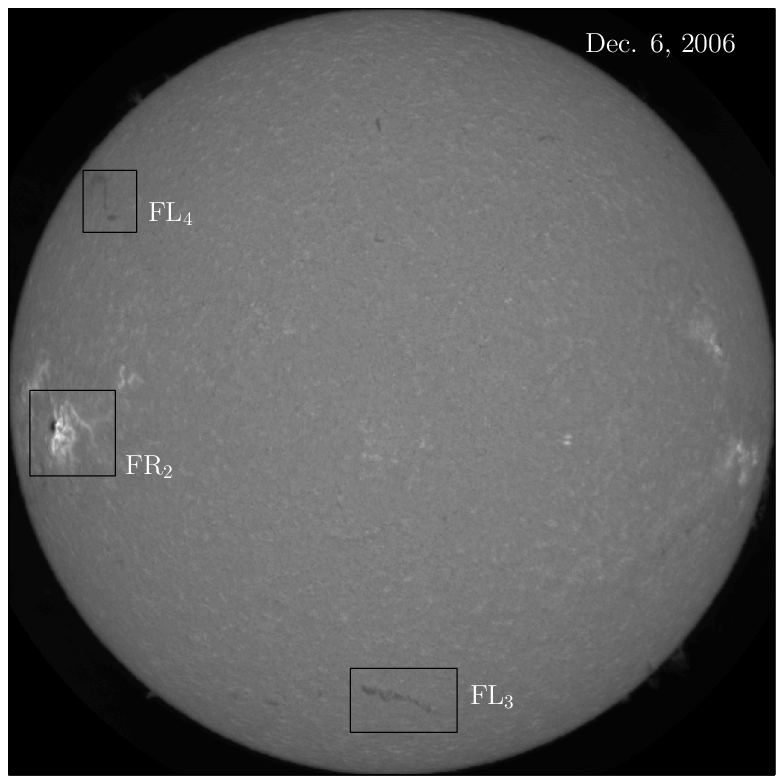}} %{fd_6_2006.eps}}
  \vspace{-.06\textwidth}
  \centerline{\hspace{.07\textwidth}\bf\color{white}{(a)}\hspace{.42\textwidth}\bf\color{white}{(b)}\hfill}
  \vspace{.02\textwidth}
  \caption{Snapshots (pre-flare) in H$\alpha$ intensity from the ISOON telescope for the (a)  2005 M-class flare and the (b) 2006 X-class flare. Three regions of interest are marked for each event, where ``FR'' denotes a flaring region and ``FL'' denotes a filament. See online version for movies of the intensity and Doppler time series. You can view all movies online at \href{http://astronomy.nmsu.edu/jasonj/Halpha\_movies/}{http://astronomy.nmsu.edu/jasonj/Halpha\_movies/}.}
  \label{fig:regions}
\end{figure}

We study two flare events observed with the United States Air Force/National Solar Observatory Improved Solar Observing Optical Network (ISOON) telescope located at Sacramento Peak, NM \citep{neidig1998}. Full-disk solar images ($2048\times 2048$ pixels at 1.1'' sampling)  at 1-minute cadence are observed in the $6563\AA$ H$\alpha$ line center with a $0.1\AA$ bandpass filter. Corresponding Doppler maps are also utilized, and have been constructed by the following procedure: 14 images across 14 wavelengths centered around the spectral line were gathered on a day in 2004 when there was little activity on the Sun. After calibrating the images for intensity corrections and flat-fields, spectral line profiles for each 1.1 arcsecond pixel were obtained, separately. From the center-of-gravity (spectral line center) for each spectral line, the intensity-difference at  $\pm 0.4\AA$, and a Doppler shift (line-center for each pixel, mean line-center for entire image) was measured. By plotting all the points of Doppler shift versus the intensity difference, a characteristic relationship is found. For any day of observing the intensity differences, a conversion of that measure to Doppler shift is computed (assuming that the calibration holds true).  In the flares and filaments considered here the H$\alpha$ line wings remain in absorption \citep{bala2004}.

A solar flare of x-ray strength  M8.0 occurred on May 13, 2005 around 16:13UT in active region NOAA 10759.  It was a classic two-ribbon flare that also corresponded to a very large CME that delivered the most intense geomagnetic activity of 2005 \citep{bisi2010}. Various characteristics of this well-known event have been studied \citep{yurchyshyn2006,liu2007,kazachenko2009}, although this is the first time oscillations in the active region and the nearby filaments have been detected.  The second event occured on December 6, 2006,  where an X-class flare erupted at about 18:40UT in NOAA 10930. For both cases, ISOON provided high-quality intensity and Doppler data from about 4 hours before until 4 hours after flare initiation. Pre-flare snapshots of full-disk H$\alpha$ intensity images for each data set are shown in  Fig.~\ref{fig:regions}. The flaring active regions (bright in H$\alpha$), as well as several quiescent filaments (dark in H$\alpha$) that this investigation targets  are outlined in boxes.

One important difference between these two flares is that the 2006 event produced a prominent Moreton wave \citep{gilbert2008,bala2010},  a fast-moving chromospheric disturbance. This particular wave propagated from the active region and interacted with several filaments on the solar disk, some quite far from the excitation location. Filament-wave interactions in this flare were studied in detail by \citet{gilbert2008}. A second important difference is the active region responsible for the 2006 event is located  close to the limb, where strong (horizontal) velocity signals should be apparent in the line-of-sight Doppler observations.

\section{Data Analysis}
\label{ana}

Studies of oscillations in filaments in the past  have used velocity time series maps from both wings of a spectral line and determine when the filament transitions from red- to blueshifted, giving an estimation of the period \citep[for example,][]{okamoto2004,gilbert2008}. Frequency content has also been derived by spatially averaging  a set of pixels in the region of interest to produce a proxy 1D ``light curve,'' and then using  wavelet techniques  \citep{jain1998,maurya2008,pinter2008}.

We demonstrate  a simple and  powerful way of visualizing these types of data, that preserves both frequency and spatio-temporal information. We generate what we call ``\emph{frequency-filtered amplitude movies}'' (FFAMs), with a method similar to what would be described as 3D wavelet analysis.

To produce the FFAMs, consider a general data cube  (we will be working with intensity and  Doppler velocity) of 2 spatial  and 1 temporal dimensions, $f(x,y,t)$. Assume there are $N$ time steps (images) available. For each time step $t_i$, the following steps are carried out:

\begin{enumerate}
\item Filter the data in the time domain over a segment of length $T$, which in this study is always taken to be 60 minutes, thus comprising $N_T=60$ images. The 60-min filter  tapers off quickly, but smoothly to zero at each ``edge.'' In other words, only the subsequent $N_T$ images after the current time step  are considered.  The filtering is represented as
  \begin{equation}
    \overline{f}_i(x,y,t; T)= F_1(t; [t_i, t_i + T])\cdot f(x,y,t),
  \end{equation}
where the brackets in the second argument of the filter $F_1$ denote the range  where it is nonzero and equal to unity. The filtered data $\overline{f}_i$ are  labeled by the particular starting time step $i$ and the segment length $T$.

\item Compute the temporal power spectrum of that segment
  \begin{equation}
    P_i(x,y,\nu;T) =  \left|\sum_{i=1}^{N_T} \,\overline{f}_i\exp(2\pi \ii\nu t_i)      \right|^2,
  \end{equation}
where $\nu$ is the cyclic frequency. Only the data over the segment $T$ will contribute to the power.  We subsequently convert this power spectrum  to an amplitude spectrum $A_i(x,y,\nu; T)$ to get units of velocity when using Doppler velocity data.

\item  Apply a  second filter  $F_2$ in frequency to the amplitude spectrum, which has a central  frequency $\nu_0$ and a bandwidth of $\Delta\nu$, and  compute the  average over the frequencies in that particular band:
\begin{equation}
 \overline{A}_i (x,y; \nu_0, \Delta\nu,T) = \langle\,   F_2(\nu;\nu_0\pm\Delta\nu/2) \, A_i(x,y,\nu; T)\, \rangle_{\nu} ,
\end{equation}
where the $\langle\ldots\rangle_\nu$ operator indicates an average over the values of frequency in the range $\nu_0\pm\Delta\nu/2$. The amplitude quantity $\overline{A}_i$ is a two-dimensional image labeled by  central frequency,  bandwidth, and segment length. 
\item Continue  for all $i\leq N-N_T$ (the final $T$ minutes  are not computed to avoid the ``edge'').  The final data product, ${\rm FFAM}=\sum_i \overline{A}_i$, is a three-dimensional quantity in $x, y, t$. It can be visualized   as a movie (see online examples).
\end{enumerate}

We  consider frequencies in 1~mHz-wide bands, centered on the values $\nu_0=1$, 3, 5, and 7~mHz. The temporal Nyquist frequency for the ISOON cadence is 8.3~mHz. The data sets in this study are approximately 420 minutes in length. Since segments from which we compute the amplitude spectrum  are only 60~min in length, the frequency sampling is $\id\nu\sim 0.28$~mHz. Therefore, the average over each 1~mHz band  uses about 3-4 images. We have done experiments with shorter and longer $T$, as well as with different bandwidths and central frequencies, and the results are qualitatively similar to  those shown in \S~\ref{res}.

%In summary, the new algorithm described above can be thought of as a type of 3D wavelet, or so-called ``short-term'' Fourier transform analysis. For a 3D cube, one could in principle compute the time-frequency wavelet spectrum of each pixel, average the relevant frequency bands, and build such an amplitude cube in a similar way. Computationally, the method described here is simpler and more efficient. 

\section{Results}
\label{res}

We compute and study FFAMs  of the oscillatory dynamics associated with the ISOON flaring events at different frequencies, and discuss the active region oscillations and the filament oscillations separately.

%Only snapshot figures of the FFAMs are shown in the print version of this article. The reader is encouraged to view the full movies in the online version.

\subsection{Oscillations in flare regions}

\begin{figure}
  \centering
    \centerline{\includegraphics[width=\textwidth,clip=true]{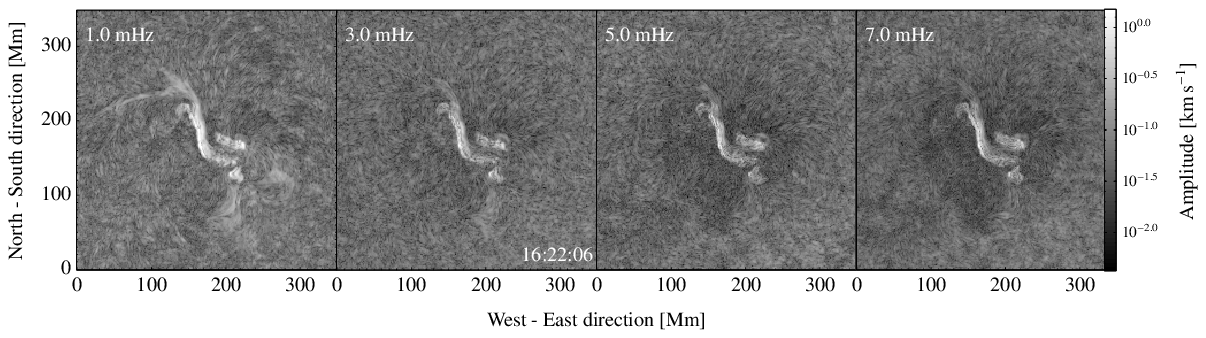}} %{fr1_2005.eps}} % frame 171
    \vspace{-.11\textwidth}
    \centerline{\bf\hspace{.07\textwidth}\Large\color{black}{FR$_1$}\hfill}
    \vspace{.07\textwidth}
    \centerline{\includegraphics[width=\textwidth,clip=true]{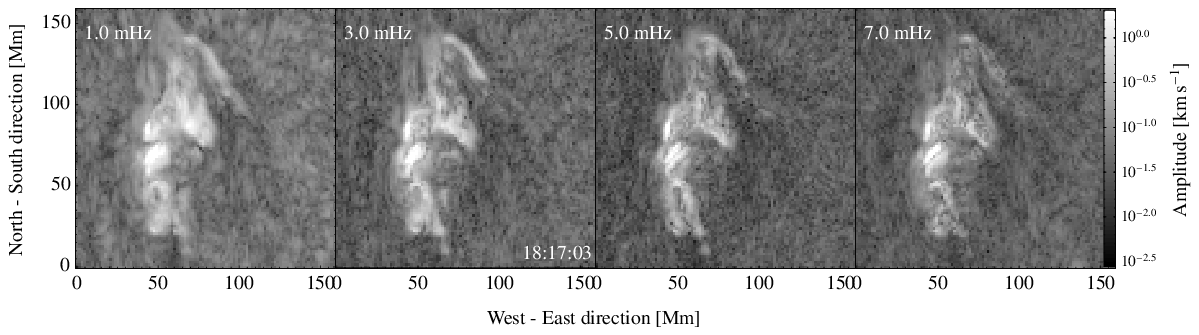}} %{fr2_2006.eps}} % frame 192
    \vspace{-.11\textwidth}
    \centerline{\bf\hspace{.07\textwidth}\Large\color{black}{FR$_2$}\hfill}
    \vspace{.05\textwidth}
    \caption{The two flaring regions. Each panel is a snapshot of an FFAM, showing the logarithm of the amplitude computed from 60~min of H$\alpha$ velocity data for the 1, 3, 5, and 7~mHz frequency bands, from left to right. The top (bottom) row shows the amplitude in the flare region FR$_1$ (FR$_2$)  from  segments centered at about 9 and 23 minutes after flare initiation, respectively. The color scale  is clipped at 0.5 of the maximum velocity amplitude to emphasize smaller scale features. See online version for full FFAMs of both intensity and velocity for each frequency and flare event.}
    \label{flares}
  \end{figure}

The active regions in which the flares occur in the 2005 and 2006 events show  oscillatory features over the entire frequency range studied here. The likelihood of a strong contribution from the underlying photospheric acoustic $p$ modes can be ruled out for several reasons.  The dramatic increase in amplitude occurs only once the flare erupts. If the  $p$ modes were  responsible, their effects would be present over the entire data sequence. Secondly, the frequency range considered here is mostly below the photopsheric acoustic-cutoff frequency ($\sim 5.3$~mHz). The filter bandpass of ISOON is also quite narrow, much more so than the instrument used by  \citet{harvey1993}, where both 5- and 3-minutes oscillations were seen in H$\alpha$. We are confident that  the effects studied here are confined mostly to the chromosphere.

Figure~\ref{flares}  shows oscillations as a function of frequency in one snapshot of the FFAM from velocity data for flare regions FR$_1$ and FR$_2$.  The central time of the 60-min segment used in the computation is provided. The figures represents the slice at  the maximal amplitude of the time series. For FR$_1$, the UT~16:22 moment is about 9 minutes after the flare erupted. Thus, there is both pre- and post-flare signal  in the derived amplitude. For FR$_2$, the 18:17 slice is about 23 minutes before the flare erupted, thus there is mostly pre-flare  signal  involved in the computation.

It is evident from the figure that the flares ``excite'' all frequencies from about 0.5 to 7.5~mHz to detectable amplitudes in velocity (note the figure is on a logarithmic scale). We are unaware of any previous studies that demonstrate this.  The velocities in the 1~mHz band reach about  $3\,{\rm km\,s^{-1}}$ and  $4\,{\rm km\,s^{-1}}$  at peak amplitude for 2005 and 2006 data, respectively, and are usually reduced at higher frequencies. The stronger amplitudes in FR$_2$ could be due to the fact that it is near the limb, thus implying stronger horizontal motions. Interestingly, the 2006 flare's maximal amplitude is in the 3 mHz band, at about $5\,{\rm km\,s^{-1}}$. In intensity the higher frequencies near each flare contribute much less oscillatory power, as is discussed further in Sec.~\ref{sec:compare}.

%Well before the flaring events the FFAM slices  show very little oscillatory behavior.  

 Once the flare erupts, the excess oscillatory amplitude  lasts for about 100 minutes after the peak in the FFAM, although this strongly depends on the location near the ribbons. Some regions are damped much faster or more strongly than others.  Due to the 60~min time-averaging involved in creating the FFAMs, it is not convenient to determine the precise duration of the increase in oscillation amplitude, as the time resolution is insufficient to give certainty to this number.  More spatial structure is seen  in the 1~mHz band, especially in the plage area of the active regions, and  the two flare ribbons in FR$_1$ are somewhat visible in the oscillation amplitudes, as is the sunspot on the northern edge. FFAMs were computed using the ISOON H$\alpha$ line core (intensity) data as well, and the same general spatial characteristics  are seen, except for weaker power at high frequencies.

\subsection{Low-frequency oscillations in filaments}

\begin{figure}
  \centering
  \centerline{\includegraphics[width=\textwidth,clip=true]{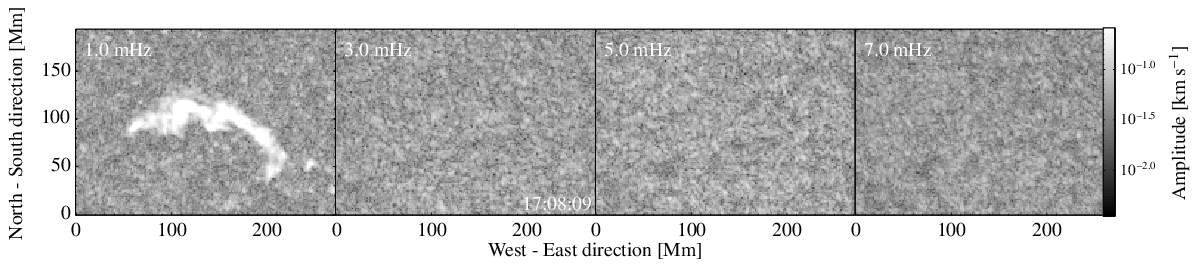}} %{fl1_2005.eps}} % 215
  \vspace{-.08\textwidth}
    \centerline{\bf\hspace{.07\textwidth}\Large\color{black}{FL$_1$}\hfill}
    \vspace{.05\textwidth}

    \centerline{\includegraphics[width=\textwidth,clip=true]{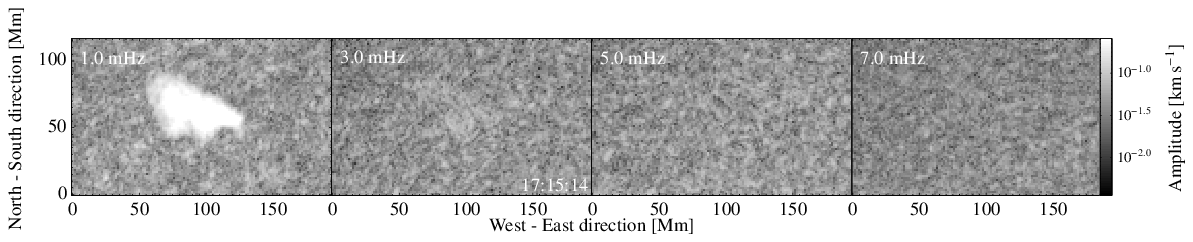}} %{fl2_2005.eps}}  % 222
    \vspace{-.08\textwidth}
    \centerline{\bf\hspace{.07\textwidth}\Large\color{black}{FL$_2$}\hfill}
    \vspace{.05\textwidth}
    \caption{Filaments during the May 13, 2005 flare. Each panel is a snapshot of an FFAM, showing the logarithm of the amplitude computed from 60~min of H$\alpha$ velocity data for  the 1, 3, 5, and 7~mHz frequency bands, from left to right. The top (bottom) row shows the amplitude in two quiescent filament regions FL$_1$  (FL$_2$) for each frequency passband, about 55 and 62 minutes after flare eruption, respectively. The color scale  is clipped at 0.5 of the maximum velocity amplitude. See online version for full FFAMs.}
  \label{2005}
\end{figure}

Throughout these two data sets the solar disk  was littered with quiescent filaments. We identify and analyze two of them for each event, denoted by ${\rm FL_{1-4}}$. The FFAMs  show a very dramatic ``lighting up'' (a rapid increase in oscillation amplitude) of the filaments well after the distant flares erupt, although each event has somewhat different characteristics.

Figure~\ref{2005} shows snapshots from the FFAM of  filaments FL$_1$  and FL$_2$ in the 2005 time series. The images  are those where the filament oscillation amplitude peaks in the FFAM, corresponding to the  time segment centered at 55 and 62 minutes after the flare erupts for FL$_1$ and FL$_2$, respectively. One explanation of the different timings is that  the filaments are at different distances from the flare center (see Fig.~\ref{fig:regions}), and whatever activation mechanism is at work (such as a chromospheric wave) needed to travel different distances.

The noticeable difference compared to the oscillations in the flare region, is that the filament only shows strong amplitude excess in the lowest frequency bandpass of $1\pm 0.5\,{\rm mHz}$, even on the logarithmic scale. This corresponds to periods in the range from about 11 to 33 minutes. The amplitude is significantly above the background level in this passband, lasting for about 100 minutes.  The velocities of the oscillations are on the order of $0.5-1.0\,{\rm km\,s^{-1}}$. This is the first detection of oscillatory behavior in these two particular filaments.

As in previous studies \citep{tripathi2009}, one might conclude that anywhere from about 3 to 10 periods of the filament oscillation took place after it was excited. With our method, it may be too simplistic  to make such a conclusion, since there are  small-scale spatial structures that are operating on different time scales, as is clearly evident in the FFAMs. Some regions show low-frequency power for the whole duration (like the central part of the filament), while others do not (the ``ends'' of the filament).

\begin{figure}
  \centering
   
    \centerline{\includegraphics[width=\textwidth,clip=true]{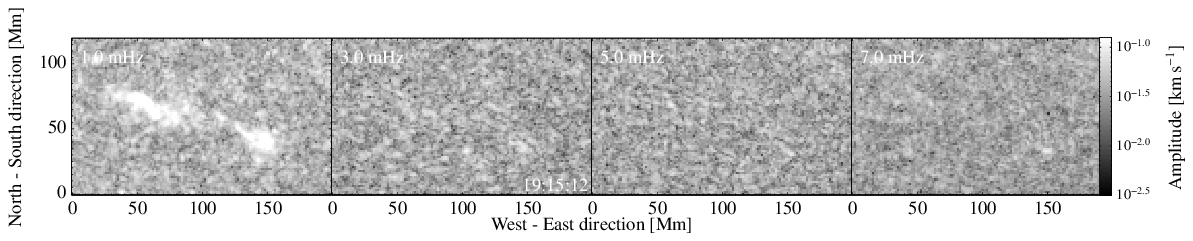}} %{fl3_2006.eps}} % 247
    \vspace{-.09\textwidth}
    \centerline{\bf\hspace{.07\textwidth}\Large\color{black}{FL$_3$}\hfill}
    \vspace{.055\textwidth}

    \centerline{\includegraphics[width=\textwidth,clip=true]{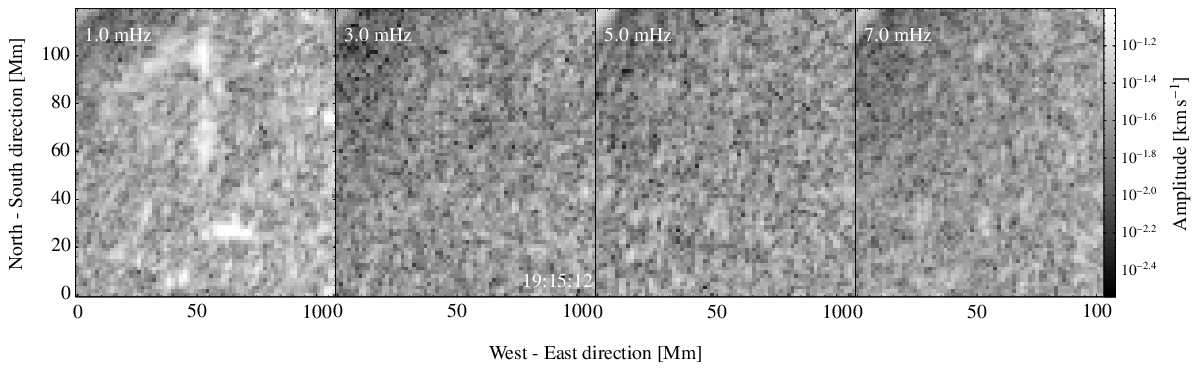}} %{fl4_2006.eps}} %247
    \vspace{-.11\textwidth}
    \centerline{\bf\hspace{.07\textwidth}\Large\color{black}{FL$_4$}\hfill}
    \vspace{.05\textwidth}
   
  \caption{Filaments during the December 6, 2006 flare. Each panel is a snapshot of an FFAM, showing the logarithm of the amplitude computed from 60~min of H$\alpha$ velocity data for  the 1, 3, 5, and 7~mHz frequency bands, from left to right. The top (bottom) row shows the amplitude in two quiescent filament regions FL$_3$  (FL$_4$) for each frequency passband, about 35 minutes after flare eruption. The color scale  is clipped at 0.5 of the maximum velocity amplitude. See online version for full FFAMs.}
  \label{2006}
\end{figure}

The 2006 quiescent filaments are shown in Fig.~\ref{2006}. We note again that a  strong Moreton wave was associated with this flare that propagated outwardly from the active region at speeds up to $1600\,{\rm km\,s^{-1}}$ \citep{bala2010}. The wave passed near both filaments and presumably activated their oscillations. The maximum values of the amplitudes in this case are  $\approx 0.2\,{\rm km\,s^{-1}}$, less than half the strength than found in  the FL$_1$ and FL$_2$ filaments. Filament FL$_3$ shows no oscillations in its center, only on its ends, which is a curious feature. Additionally, even though these two filaments are not equidistant to the flare region, the maximal oscillation amplitude occurred at approximately the same time.

That the velocity oscillation signal in the filaments is  weaker than in the 2005 event is surprising, given that  there is a clear and strong Moreton wave observed that interacts with the filaments.  Possible explanations for this discrepancy might include viewing geometry, intrinsic difference in filament properties, lower quality of data, or the overall distance of the filaments from the flare, which is larger than in the 2005 case.

%% MAKE COMPARISONS TO "OLD METHODS"
\subsection{Comparison of method to other analysis techniques}
\label{sec:compare}

\begin{figure}
  \centering
  \centerline{\includegraphics[width=\textwidth,clip=true]{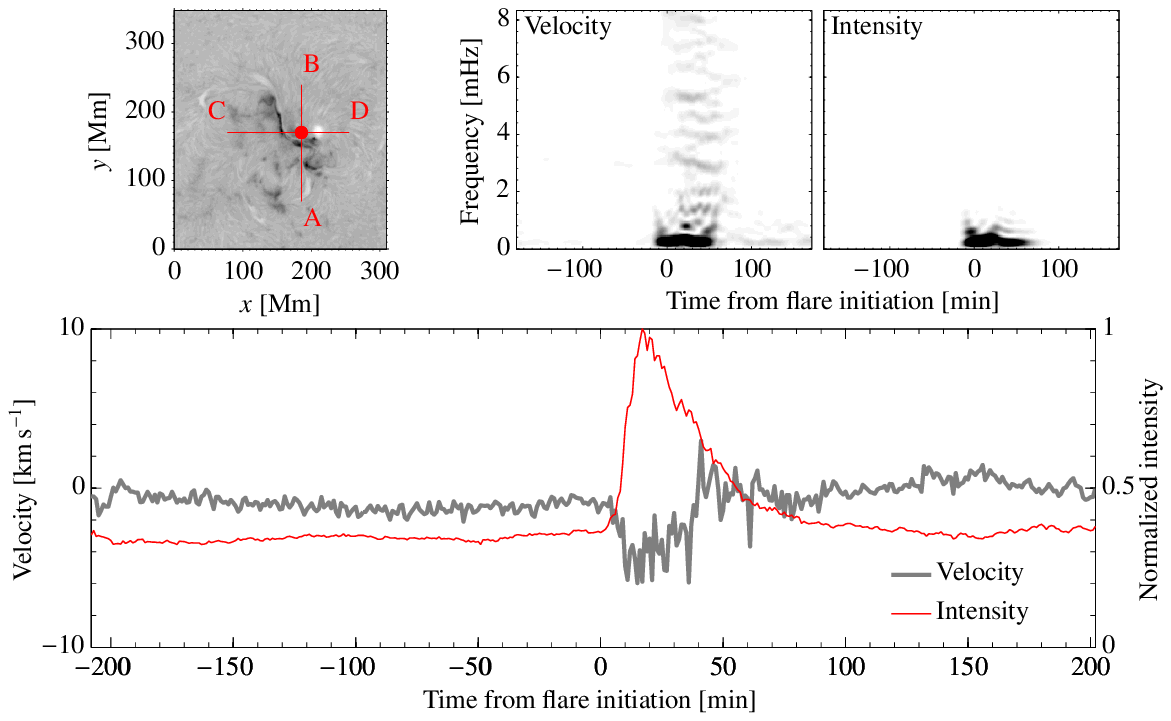}}
  \vspace{-.66\textwidth}
  \centerline{\hspace{.05\textwidth}\bf\color{black}{(a)}\hspace{.3\textwidth}\bf\color{black}{(c)}\hfill}
  \vspace{.3\textwidth}
  \centerline{\hspace{.1\textwidth}\bf{(b)}\hfill}
  \vspace{.3\textwidth}
  \caption{Analysis of flare FR$_1$ using standard methods. (a) A snapshot of intensity around the flare. The red circle indicates the pixels (radius of 10 pixels) that were averaged to compute the time series in panel (b). The horizontal and vertical lines denote slices used for the plots in Fig.~\ref{stack_fr1}. (b) Time series of normalized intensity and Doppler velocity for the set of pixels in the flare region as a function of time. (c) Wavelet power spectra of the velocity and intensity light curves from (b). The (linear) gray scale in these plots is such that black denotes higher signal, and  has been clipped to half of the maximum power to emphasize details.}
  \label{wave_fr1}
\end{figure}

\begin{figure}
  \centering
  \includegraphics[width=\textwidth,clip=true]{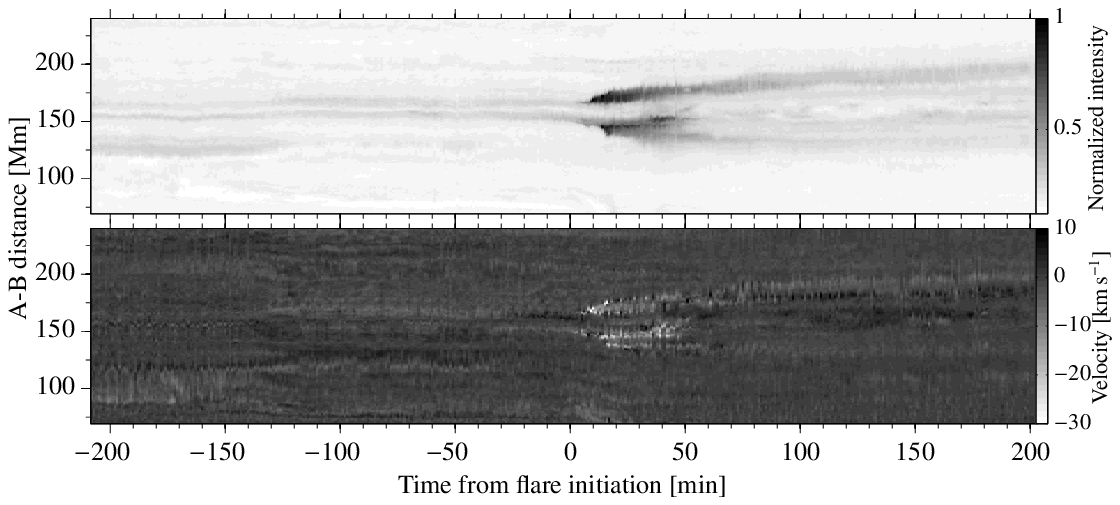}
  \includegraphics[width=\textwidth,clip=true]{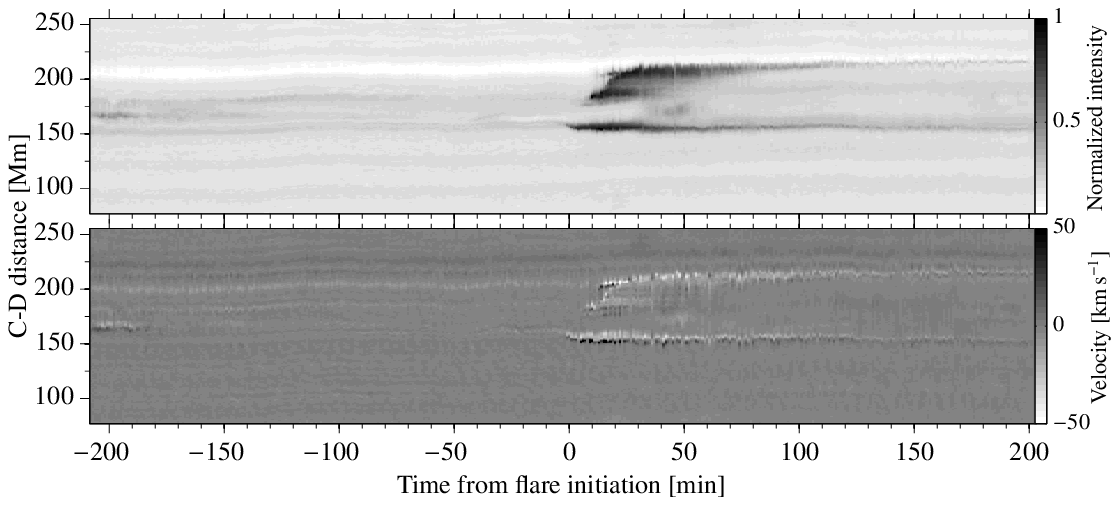}
  \caption{Stack plots of flare FR$_1$ along the slices indicated in Fig.~\ref{wave_fr1}a. The top two panels show the intensity and velocity time dependence along the A-B vertical line through the flare. The bottom two panels  show the intensity and velocity time dependence along the C-D horizontal line through the flare.}
  \label{stack_fr1}
\end{figure}

Analyzing one-dimensional time series of various regions by wavelet analysis is one technique that has become standard for the study of filament oscillations \citep[e.g.,][]{pinter2008}. Computing ``stack plots'' of the signal along a certain slice of the data as a function of time  is another \citep[e.g.,][]{li2012}. We carry out such analyses to compare and contrast the results with those of the FFAMs for the flare  FR$_1$ and filament FL$_1$ regions.  The other regions show similar behavior.

Figure~\ref{wave_fr1} shows an analysis of FR$_1$. A set of pixels near the flare ribbons was averaged and time series of intensity and velocity were obtained. The intensity light curve shows the characteristic flare impulse and exponential decay, accompanied by low-level oscillations, while the velocity shows a fluctuating signal on top of a sudden blueshift (towards observer) as the flare erupts. Other pixels in the flare-ribbon region show redshifts at flare onset \citep[for a detailed discussion of velocity profiles in broadband H$\alpha$ measurements, see][]{ichimoto1984}. In both cases the frequency content of the oscillations are difficult to quantify in the time series alone.  Wavelet analysis of this set of pixels confirms that all frequencies in velocity are activated, but that the low frequencies are strongest, as the FFAMs indicated. The strongest amplitudes last for about 80 minutes. The spectrum in intensity does not show much high-frequency power. These conclusions change somewhat as different regions are analyzed. 

Perpendicular slices through the flare ribbons passing through the center of the same set of pixels are used to produce stack plots for intensity and velocity and are given in Fig.~\ref{stack_fr1}.  From the horizontal slice one sees that the east side of the flare (closest to point C) is the first to begin oscillating, as is evident also in the FFAM. The strongest signal comes from the outer edge of the ribbons. The velocity stack plots indicate that the  east-west direction (C-D) has a higher signal-to-noise,  even though the other direction has the larger overall amplitude.

\begin{figure}
  \centering
  \centerline{\includegraphics[width=\textwidth,clip=true]{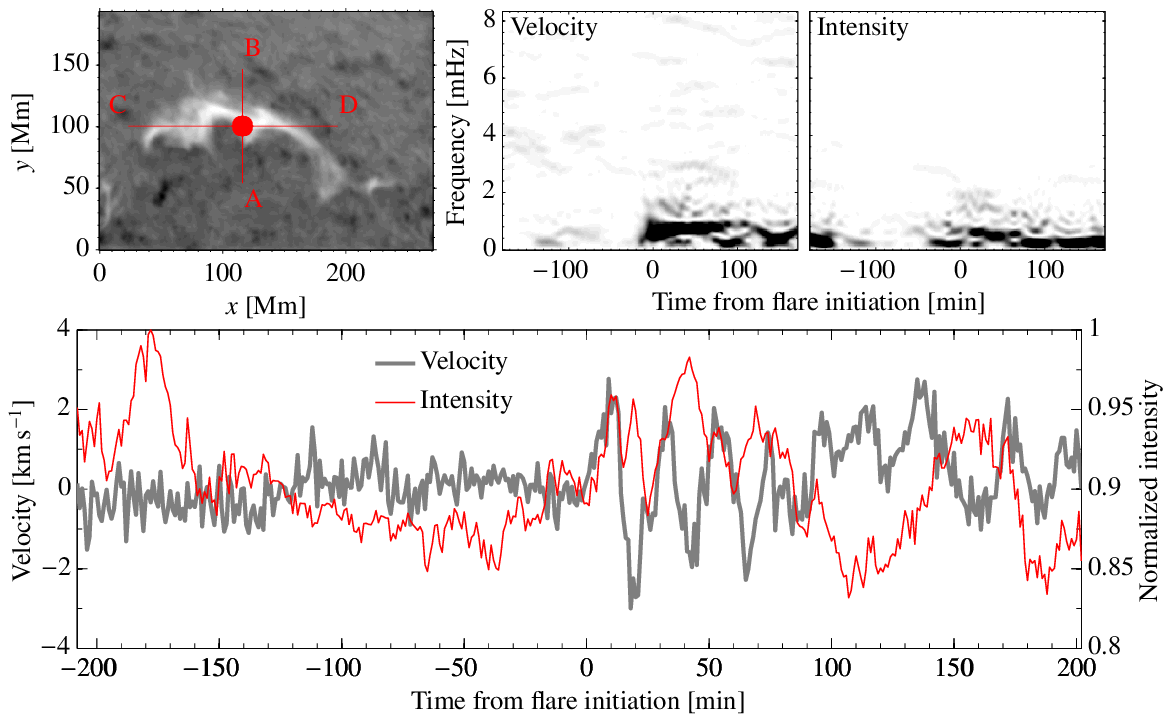}}
  \vspace{-.67\textwidth}
  \centerline{\hspace{.05\textwidth}\bf\color{black}{(a)}\hspace{.3\textwidth}\bf\color{black}{(c)}\hfill}
  \vspace{.3\textwidth}
  \centerline{\hspace{.07\textwidth}\bf{(b)}\hfill}
  \vspace{.3\textwidth}
  \caption{Analysis of filament FL$_1$ using standard methods. (a) A snapshot of intensity around the filament. The red circle indicates the pixels (radius of 10 pixels) that were averaged to compute the time series in panel (b). The horizontal and vertical lines denote slices used for the plots in Fig.~\ref{stack_fl1}. (b) Time series of normalized intensity and Doppler velocity for the set of pixels in the filament region as a function of time. (c) Wavelet power spectra of the velocity and intensity light curves from (b). The (linear) gray scale in these plots is such that black denotes higher signal, and  has been clipped to half of the maximum power to emphasize details.}
  \label{wave_fl1}
\end{figure}

\begin{figure}
  \centering
  \includegraphics[width=\textwidth,clip=true]{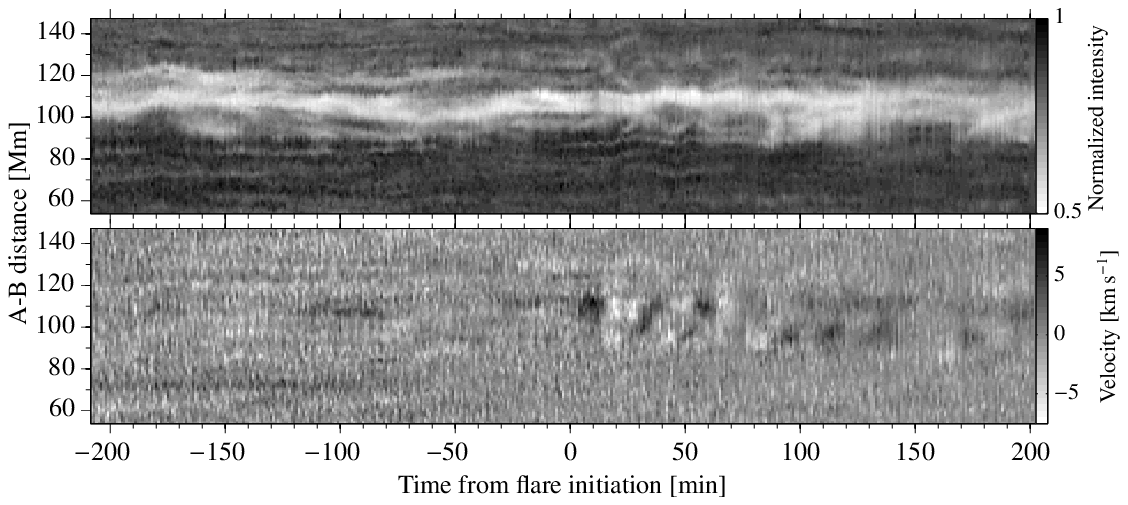}
  \includegraphics[width=\textwidth,clip=true]{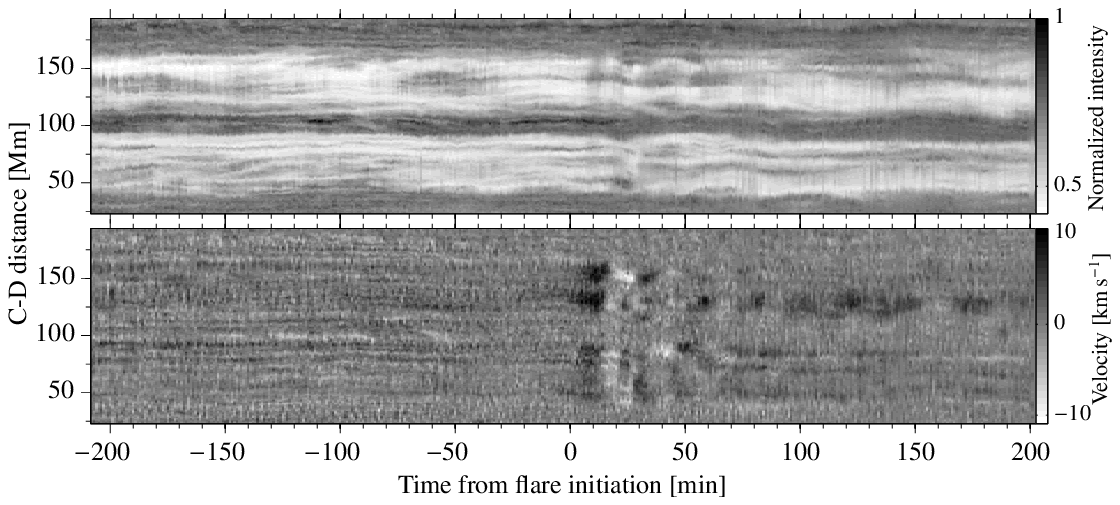}
  \caption{Stack plots of filament FL$_1$ along the slices indicated in Fig.~\ref{wave_fl1}a. The top two panels show the intensity and velocity time dependence along the A-B vertical line through the filament. The bottom two panels  show the intensity and velocity time dependence along the C-D horizontal line through the filament.}
  \label{stack_fl1}
\end{figure}

Figure~\ref{wave_fl1} shows a similar analysis for the filament FL$_1$. A set of pixels near the center of the filament that showed strong oscillations in velocity were chosen and averaged to give the time series. There are 6-7 long-period oscillations ($\sim 20~{\rm min}$) at amplitudes of about $1~{\rm km\,s^{-1}}$ in this region, and these show up clearly in the wavelet spectra. In intensity, the variations in the time series are less well defined, but the wavelet spectra do reveal the low-frequency nature of the oscillations. Contrasting with  the chosen flaring region, the strong amplitudes near the filament persist for well over 100 minutes. The stack plots in Fig.~\ref{stack_fl1} show slices through the filament along perpendicular directions. Again, the global view of the intensity slice is not particularly useful for studying any variations, while the velocity plots clearly show the filament being triggered by the flare and its subsequent oscillations. The motion  appears to be primarily in the east-west longitudinal direction.

The use of these standard techniques are indeed very  useful for studying these phenomena, especially when coupled with an analysis using FFAMs. However, the choice of pixels and slice coordinates to produce Figs.~\ref{wave_fr1}-\ref{stack_fl1} requires a lot of trial and error, and the inferences vary strongly from region to region. This could be expedited by first using the FFAMs to choose suitable regions that show interesting behavior, and then looking into more detail with the other methods. It is interesting that the stack plots in intensity for the filament do not strongly show the flare eruption, at least along these particular slices, while the intensity FFAMs and wavelets clearly do.

%%% END COMPARISON

\subsection{Summary of results}

\begin{figure}
  \centering
  \centerline{
    \includegraphics[width=.47\textwidth,clip=true]{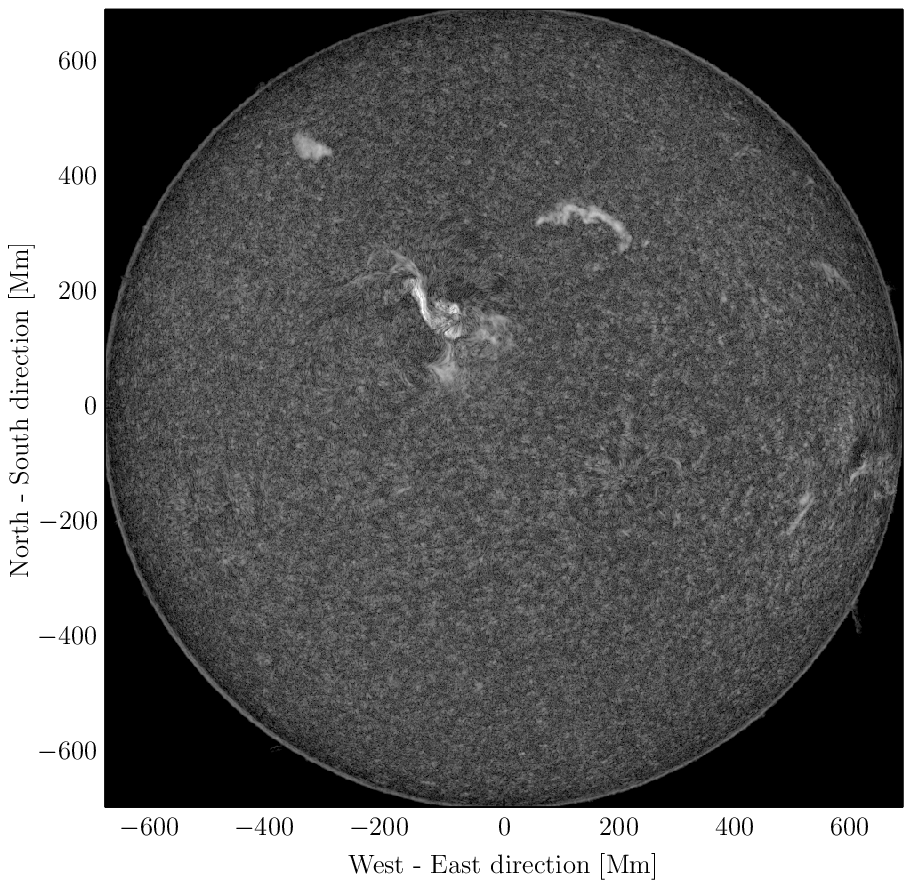}  %{fd_1mhz_2005.eps} % frames 193
    \includegraphics[width=.47\textwidth,clip=true]{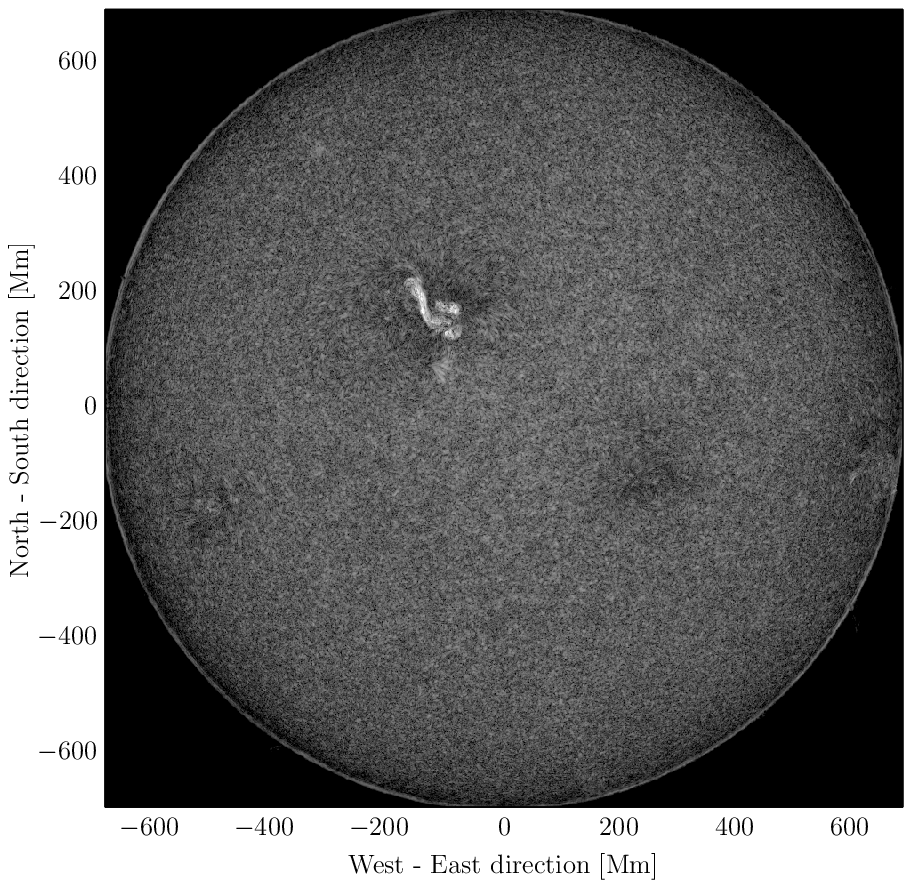}} %{fd_3mhz_2005.eps}}
  \vspace{-.1\textwidth}
  \centerline{\hspace{.1\textwidth}\bf\color{white}{(a)}\hspace{.45\textwidth}\bf\color{white}{(b)}\hfill}
  \vspace{.02\textwidth} 
  \centerline{ 
    \includegraphics[width=.47\textwidth,clip=true]{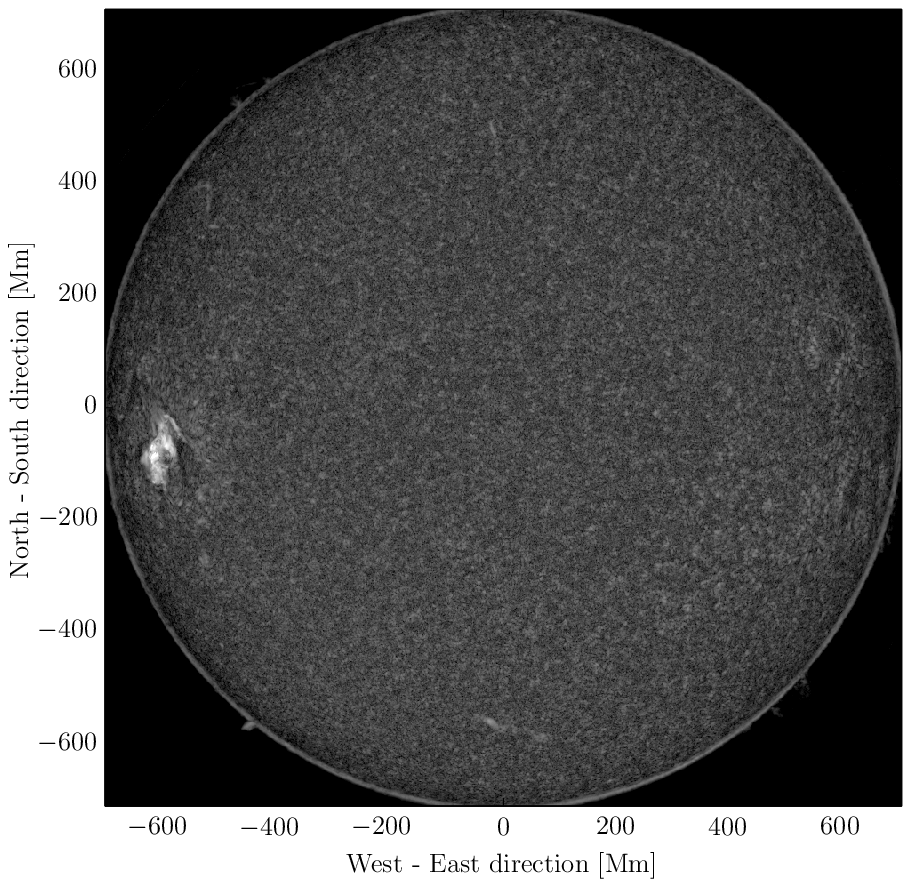}  %{fd_1mhz_2006.eps} % frames 220
    \includegraphics[width=.47\textwidth,clip=true]{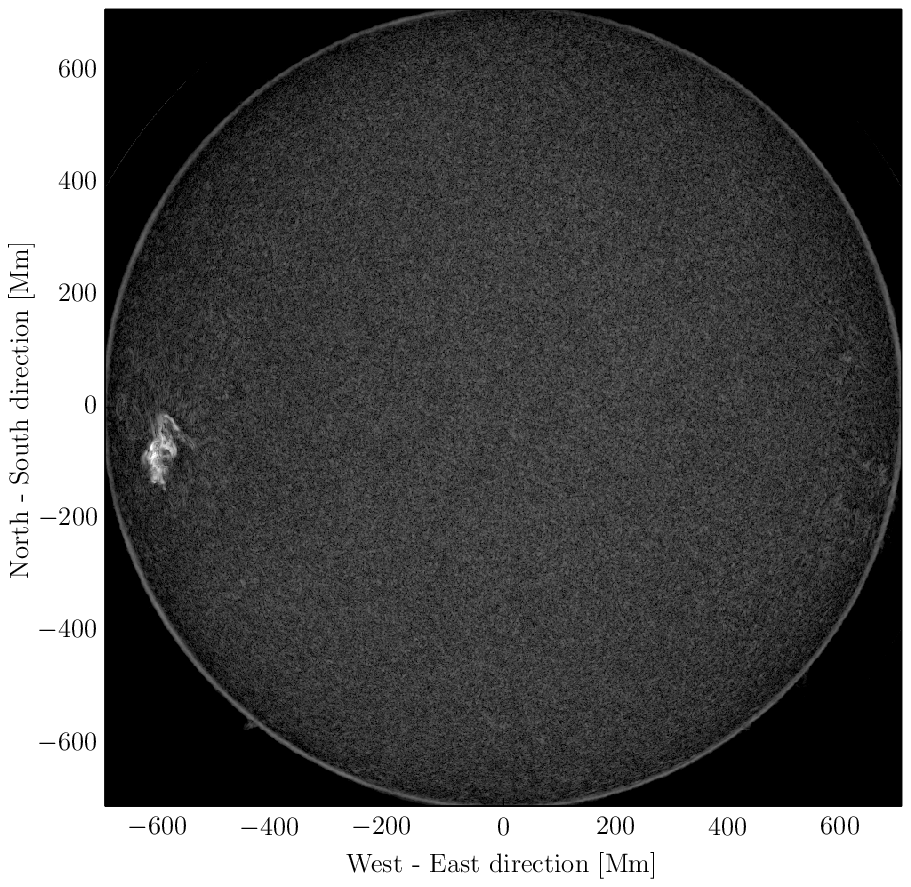}} %{fd_3mhz_2006.eps}}
  \vspace{-.1\textwidth}
  \centerline{\hspace{.1\textwidth}\bf\color{white}{(c)}\hspace{.45\textwidth}\bf\color{white}{(d)}\hfill}
  \vspace{.05\textwidth}
  \caption{Snapshots of full-disk velocity FFAMs around peak flare intensity. Panels (a) and (b) are the 1~mHz and 3~mHz bands for the 2005 flare, respectively. Panels (c) and (d) are the  1~mHz and 3~mHz bands for the 2006 flare, respectively. All panels are plotted on the same gray scale. See the full-disk FFAMs online.}
  \label{fig:fd_bands}
\end{figure}

To summarize the analyses,  we observe the fascinating occurence of rapid communication at large distances across the solar disk. In the 2005 event, strong amplitude increases on the far western solar limb are observed, very far from the flare area in the eastern hemisphere. Indeed, Fig.~\ref{fig:fd_bands} shows full-disk snapshots of velocity FFAMs in the two lowest bandpasses near the peak flaring times. The filamentary amplitudes are evident everywhere at low frequency, and rather absent at 3~mHz and higher.  For the 2006 flare, the 1~mHz band shows weak excess acoustic power just outside the flare region, and  one barely discerns the brightenings in  filaments FL$_3$ and FL$_4$ in this global picture.

A comparison with other  large-amplitude filament oscillations as reviewed by \citet{tripathi2009} and \citet{arregui2012} is useful. Other published studies find oscillations with minimum amplitudes near $20\,{\rm km\,s^{-1}}$.  \citet{gilbert2008} find in  line-of-sight velocity images from the Polarimeter for Inner Coronal Studies (PICS) H$\alpha$ instrument that the  FL$_3$ filament oscillates at amplitudes up to $41\,{\rm km\,s^{-1}}$. We find amplitudes of a few   ${\rm km\,s^{-1}}$ in the flare regions themselves, but less than $1\,{\rm km\,s^{-1}}$ in the filaments at low frequency. In the raw Doppler signal from ISOON we find maximum velocities of the FL$_3$ filament of several ${\rm km\,s^{-1}}$. 

While we do observe features that approach several tens of  ${\rm km\,s^{-1}}$ in the line-of-sight velocity data, particularly in the flare (see Fig.~\ref{stack_fr1}, for example), the coherent variations over an extended time period do not possess such amplitudes. We attribute the discrepancy with  other values in the literature to the fact that the FFAMs select out the persistent variations over longer time scales (in the computations here, at least 60~min), and thus it is a matter of effective time resolution.  Differences in instrumentation likely have some effects too.

For filament oscillation periods, studies have found a range from about 6 to 150 minutes \citep{tripathi2009}. The periods found here are on the lower end of those values, but still consistent with them. One might argue that small-amplitude oscillations are a suitable classification for these events, which tend to have a few ${\rm km\,s^{-1}}$  amplitudes, but can also have very short periods on the order of seconds to a few minutes, which we are unable to observe.

\citet{gilbert2008} have already studied both FL$_3$ and FL$_4$ using different data and standard methods. For FL$_3$, they find 3~hr of oscillatory motions, somewhat longer than found here by about 30 minutes. They also find  periods from  18-39~min. For FL$_4$ they find the duration of oscillations to be only about 17 minutes, whereas we observe increased power (at 1~mHz) for more than  1~hr. The different diagnostics as a function of frequency is an advantage of the method presented here.

\section{Discussion}
\label{dis}

This work is important for two  reasons: (1) we have presented a simple algorithm for both visualizing and studying dynamic phenomena in 2+1 dimensions with differing frequency content; (2) we have used this technique to analyze two different events that exhibit low-frequency filament oscillations excited by a distant flare. We observe robust oscillations in the active regions where the flares erupt over a broad range of frequencies.  Photospheric modes can be ruled out as the source  since the oscillation amplitude only increases after the flare initiates.

Strangely,  the weaker  filament oscillations  are observed in the system  that interacted with a  strong  Moreton wave, which can trigger of such oscillations, but  it is  possible that line-of-sight effects or wave propagation direction  can be involved in these differences. The velocities found in the FFAM analysis do not fit into the large- and small-amplitude oscillation classifications. It might be useful to have sub-categories of such events that show oscillating strengths as a function of frequency. We have been able to confirm  the work of \citet{gilbert2008} who showed that the FR$_4$ filament is a lower-amplitude oscillator than the other filament in the December 2006 flare event.

The spatial features of the velocity power in the different frequency bands show interesting dynamics, particularly for the filaments. We observe different sections of the filaments fluctuating over various timescales, indicating that either the damping of the oscillations is spatial dependent or the excitation  is not constant over the entire length of the filaments, or some combination of both effects. More detailed analysis of this is underway.

There is both new and complementary information when comparing the original full-disk time series and the FFAMs (the movies associated with Fig.~\ref{fig:regions} and the FFAMs of Fig.~\ref{fig:fd_bands}). For example, for the 2005 event, in the raw intensity time series one clearly sees the reaction of some of the far-away filaments after flare eruption, but the FFAM indicates that some filaments are actually not affected by the (unknown) trigger mechanism, and also directly provides the oscillation frequencies of those which are. In the 2006 flare the Moreton wave is easily visible in both intensity and Doppler data, yet the response of the filaments is rather muted when viewed in the FFAM, suggesting that the oscillation activation was not as effective. Certainly more events need to be treated in such a systematic fashion to understand if these are common occurences or not.

We finally point out that at the photosphere, active regions typically show an enhancement of acoustic power at \textit{higher} frequencies, known as acoustic halos \citep{braun1992}. This is quite different than the case here  where we see  higher amplitudes in the lower-frequency bandpasses. How the frequency content of waves changes through the different layers of the atmosphere could help in understanding the local environment parameters that are quite difficult to directly observe, such as magnetic field and density.  EUV waves can also presumably interact with filaments, although for the two events studied here there were no EUV data that could be used to  identify any such contribution. In the future this method will be used to study a range of various chromospheric data.

%Therefore, after some further refinements, this method will be applied to other interesting active regions, as well as to other  datasets such as the various passbands of the Atmospheric Imaging Assembly  (AIA) instrument \citep{lemen2012} onboard the Solar Dynamics Observatory.

\acknowledgements

We acknowledge fruitful conversations with Stuart Jefferies, Michael Kirk, Frank Hill, and R.T.~James McAteer. This work was supported in part by an NSF PAARE award AST-0849986.

%\bibliographystyle{apj}
%\bibliography{myrefs}

\end{document}